\documentclass[twoside,journal]{IEEEtran}

\usepackage{graphicx}
\usepackage{fancyhdr}

\usepackage{makecell}
\usepackage{hyperref}
\usepackage{array}
\usepackage{amssymb,amsmath}
\usepackage{multicol}
\usepackage[noadjust]{cite}
\usepackage{setspace}
\usepackage{subfigure}
\usepackage{float}
\usepackage{url}
\usepackage{stfloats}
\usepackage{amsthm,pifont}
\usepackage{flushend}
\usepackage{cases,subeqnarray}
\usepackage{bm,multirow,bigstrut}
\usepackage{textcomp}
\usepackage{latexsym}
\usepackage{booktabs}
\usepackage{xcolor}
\usepackage{mathtools}
\usepackage{dsfont}
\usepackage{extarrows}
\usepackage{epsfig}
\usepackage{epstopdf}
\usepackage{algorithm}
\usepackage{algpseudocode}
\usepackage{booktabs}
\usepackage{tabularx}
\usepackage{array}
\usepackage[table]{xcolor}

\definecolor{wirelessblue}{RGB}{235,244,250}
\definecolor{jepabeige}{RGB}{250,246,232}
\definecolor{intelligencegreen}{RGB}{237,247,239}

\newcolumntype{Y}{>{\raggedright\arraybackslash}X}
\fancypagestyle{labpaperfirst}{
    \fancyhf{}
    \fancyhead[L]{%
        \includegraphics[width=4.2in]{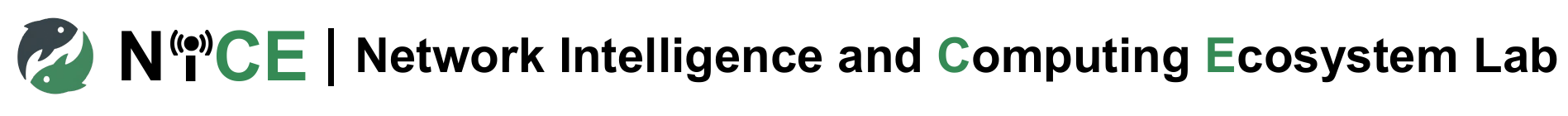}
    }

}

\theoremstyle{plain}

\begin{document}

\title{CommJEPA: When Predictive Representation Learning Meets Wireless Intelligence}
\author{Feiran You, Hongyang Du, Kang Wei, Jun Li,~\IEEEmembership{Fellow,~IEEE},  Abbas Jamalipour,~\IEEEmembership{Fellow,~IEEE}
\thanks{F. You and H. Du are with the Department of Electrical and Computer Engineering, The University of Hong Kong, Pok Fu Lam, Hong Kong SAR, China (email: youfr@hku.hk, duhy@hku.hk) (Corresponding Author: Hongyang Du).}
\thanks{Kang Wei is with the School of Cyber Science and Engineering, Southeast University, Nanjing 211189, China (e-mail: kang.wei@seu.edu.cn)}
\thanks{J. Li is with the National Mobile Communications Research Laboratory, Southeast University, Nanjing 210096, China (e-mail:  jun.li@seu.edu.cn)}
\thanks{{\color{black}A. Jamalipour is with the School of Electrical and Computer Engineering, University of Sydney, Australia, and with the Graduate School of Information Sciences, Tohoku University, Japan (e-mail: a.jamalipour@ieee.org).}}
}
\maketitle
\thispagestyle{labpaperfirst}
\vspace{-1cm}

\begin{abstract}
Artificial intelligence (AI) is becoming an integral part of wireless systems, yet much of today's wireless AI remains task-specific and reactive, with models typically trained to map current observations to predefined outputs or reconstruct observed data. While effective for individual tasks, such approaches provide limited support for learning what is predictable, reusable, and transferable in dynamic wireless environments. Joint-Embedding Predictive Architecture (JEPA) offers a different perspective by learning predictive representations in latent space that capture informative structure in the observations. In this article, we introduce \emph{Communication JEPA} (\emph{CommJEPA}) as a communication-oriented framework for predictive representation learning. The central idea is bidirectional. JEPA can equip wireless systems with predictive representations for capturing latent channel and environmental dynamics, integrating heterogeneous observations, adapting to changing conditions, and supporting future decisions. Conversely, the physical structure, partial observability, multiple time scales, action-dependent evolution, and resource constraints of wireless systems motivate new communication-aware JEPA designs. We develop this mutual perspective, organize the resulting design space, and discuss representative JEPA-based wireless instantiations for prediction and control. We further highlight open challenges toward scalable, interactive, and predictive wireless intelligence.
\end{abstract}

\section{Introduction}

Artificial intelligence (AI) has become an important enabler of modern wireless communication systems~\cite{11155147}. Learning-based methods have been widely applied to channel estimation, beam management, sensing, resource allocation, mobility management, and network control~\cite{10930485}. More recently, advances in self-supervised learning and foundation models have shifted attention toward a broader goal: learning wireless representations that remain useful across tasks, environments, and operating conditions. This goal is increasingly important as future networks move from isolated learning modules toward more general and adaptive wireless intelligence.

Much of today's wireless AI, however, remains task-specific and largely reactive~\cite{11611638}. Supervised learning methods typically map current observations to predefined outputs, such as channel estimates, beam selections, resource-allocation decisions, or control actions, while reconstruction- and generation-based approaches focus on recovering or synthesizing observed signals~\cite{11316403}. Reinforcement learning enables sequential decision-making through reward-driven policy learning, but the resulting policies may remain closely tied to specific objectives and training environments~\cite{11248867}. Numerical optimization provides principled solutions to many wireless resource-allocation and control problems, yet its applicability can depend on tractable system models and may be constrained by computational complexity in dynamic, large-scale networks. Despite their effectiveness in individual settings, these approaches do not necessarily provide reusable representations that capture the underlying structure of wireless environments across tasks and changing conditions. This limitation is particularly important in dynamic networks, where observations offer only partial information about the environment, relevant conditions evolve over time, and communication or control actions may themselves influence subsequent system states. Wireless intelligence should therefore move beyond task-specific prediction and isolated decision-making toward representations that capture information likely to persist, evolve, or remain relevant for future decisions~\cite{jiang2026large}.

Joint-Embedding Predictive Architecture (JEPA) provides a promising learning principle for this purpose~\cite{lecun2022path,assran2023self}. JEPA learns by predicting target representations in latent space, allowing models to capture structure that remains predictable across observations, system states, and time points. In wireless systems, such structure may reflect propagation geometry, mobility patterns, environmental conditions, and network dynamics. This approach is particularly relevant because raw wireless measurements often contain rapidly varying, noisy, or task-specific components, whereas propagation characteristics, mobility trends, and other latent regularities may exhibit more persistent patterns that transfer across downstream tasks.

\begin{figure*}[t]
    \centering
    \includegraphics[width=\linewidth]{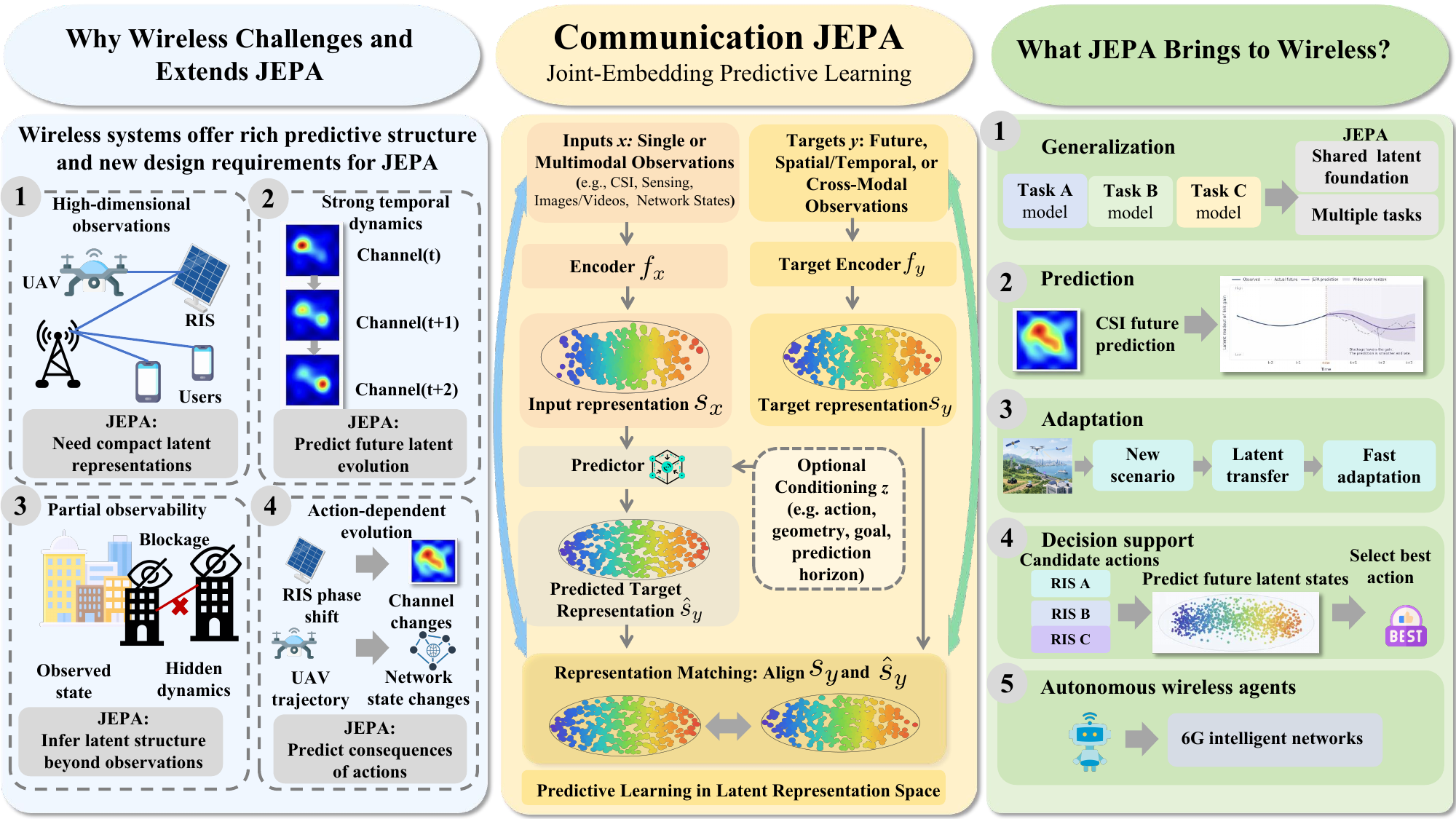}
    \caption{The mutual evolution of JEPA and wireless intelligence. CommJEPA bridges predictive representation learning and wireless systems: wireless characteristics shape what and how JEPA should predict, while JEPA provides reusable predictive representations for understanding, adaptation, and decision-making in future wireless networks.}
    \label{fig:commjepa_framework}
\end{figure*}

JEPA offers wireless systems a framework for predictive representation learning. The relationship is reciprocal, as the physical, temporal, observational, interactive, and resource constraints of wireless environments shape how this framework is designed and which information it represents and predicts. These characteristics can in turn motivate new forms of predictive representation learning. JEPA may therefore provide wireless systems with reusable and anticipatory representations, while wireless communication problems can reshape the design of JEPA itself. Motivated by this bidirectional relationship, we introduce \emph{Communication JEPA} (\emph{CommJEPA}), as illustrated in Fig.~\ref{fig:commjepa_framework}. CommJEPA provides a communication-oriented framework for designing predictive learning objectives around wireless observations and system evolution, with flexibility across neural architectures and wireless tasks. JEPA enables latent representation learning to support wireless understanding, adaptation, and decision-making, while the physical and operational characteristics of wireless systems shape the information to be represented, the targets to be predicted, and the conditions under which prediction takes place. This reciprocal relationship establishes a common perspective for advancing wireless AI from task-specific reactive systems toward reusable predictive intelligence.

In this article, we develop this perspective in three steps. 
\begin{itemize}
    \item First, we revisit joint-embedding predictive learning from a wireless communication viewpoint and define the basic principles of CommJEPA. 
    \item Second, we examine the mutual relationship between wireless systems and JEPA, showing both how predictive representations can benefit wireless intelligence and how communication-specific characteristics motivate new JEPA designs. 
    \item Third, we connect predictive representations to wireless decision-making through two complementary instantiations, including future-state prediction for decision-oriented optimization and preference-aware prediction for wireless control.

\end{itemize}
 We finally discuss the main research challenges toward physics-grounded, multimodal, interactive, generalizable, and resource-efficient predictive wireless intelligence.

\section{From JEPA to Communication JEPA}

This section develops the formulation of CommJEPA. We first introduce predictive representation learning in wireless systems and define the central principle of CommJEPA. We then explain how communication-specific structure shapes the predictive learning problem and identify the wireless tasks for which such representations are most relevant. The corresponding design dimensions are discussed in Section~III-B.

\subsection{Predictive Representation Learning for Wireless Systems}

Conventional wireless AI commonly learns task-specific mappings from observations to predefined outputs such as channel estimates, beam selections, resource allocations, or control actions~\cite{yang2025decision}. JEPA instead learns predictive representations by encoding available observations and predicting the latent representation of a related target, which may correspond to another spatial region, modality, or future state. This formulation is particularly relevant to wireless systems, where fading, mobility, blockage, hardware effects, and network states evolve across different spatial and temporal scales. Latent prediction can focus on structures that remain predictable and relevant across observations, locations, modalities, and time. The central design question is which predictive relationships should be learned so that the resulting representations preserve physically meaningful and decision-relevant information.

We define CommJEPA as a predictive representation-learning framework guided by the physical structure and decision requirements of wireless systems. Wireless dynamics, physical constraints, observability, candidate actions, and communication objectives determine which predictive structures are learned and preserved in latent space. These factors shape observation organization, prediction targets and horizons, conditioning information, and the distinctions retained in the learned representation. CommJEPA organizes future-relevant wireless information into predictive abstractions that support adaptation, optimization, and control.

The input may include physical-layer signals, channel state information (CSI), sensing measurements, positioning information, environmental observations, visual data, or network states collected across multiple sources and modalities. Prediction targets may correspond to future propagation conditions, hidden system states, cross-modal information, or action-dependent consequences. Conditioning information may incorporate temporal history, geometry, physical knowledge, candidate actions, prediction horizons, and communication objectives. Their selection depends on the predictability of the wireless process and the information required by the intended communication decision. Operational objectives can guide target and horizon selection, determine which latent distinctions receive emphasis, or directly enter the conditioning information or predictive target.

CommJEPA can support temporal, spatial, cross-modal, and action-conditioned prediction. Temporal prediction captures the evolution of latent wireless states, spatial prediction exploits dependencies across antennas, users, access points, and geographic regions, and cross-modal prediction captures relationships among radio, sensing, positioning, visual, and network-side observations. In interactive settings, candidate actions can condition the predictor to characterize system evolution under different interventions. The resulting representations can be used by downstream optimizers, reinforcement-learning policies, or controllers.

\subsection{Communication-Aware Predictive Design}

Recent work has demonstrated how JEPA can be adapted to wireless signals. WirelessJEPA, for example, organizes multi-antenna IQ samples into an antenna--time representation and introduces structured masking across antenna and temporal dimensions to learn transferable latent representations~\cite{chu2026wirelessjepamultiantennafoundationmodel}. Its predictive objective exploits the spatio-temporal structure of multi-antenna observations, showing how wireless signal organization can provide useful inductive biases for JEPA-based representation learning.

CommJEPA extends this perspective from signal organization to the formulation of the predictive learning problem. In wireless systems, propagation, observation mechanisms, protocol operation, candidate actions, and communication objectives influence which latent regularities are useful to learn. The role of these factors goes beyond choosing an input representation. They affect which states can be inferred, which future relationships are predictable, and which distinctions should be preserved for subsequent communication decisions.

Some of these issues, such as partial observability, heterogeneous observations, and action-dependent dynamics, also appear in general predictive models and world models. Their role in wireless systems is shaped by communication-specific mechanisms. Wireless states are often observed through pilots, received signals, sensing outputs, distributed measurements, and delayed feedback, linking observability to propagation geometry, measurement availability, and protocol operation. Predictability also varies across processes with different physical and operational time scales, including channel variation, mobility, blockage, traffic, and topology changes. Communication actions such as beam selection, RIS configuration, UAV movement, scheduling, and power allocation can alter subsequent connectivity, interference, and resource availability. Radio, sensing, positioning, and visual observations can further provide complementary information about the same environment. Communication and computation constraints then limit how much of this predictive information can be acquired, represented, exchanged, and processed. The corresponding design implications are developed in Section~III-B.

This perspective also clarifies the operating regime of CommJEPA. It is most relevant to wireless tasks where temporal, spatial, cross-modal, or action-dependent dependencies contain information useful for future prediction or decision-making. Mobility management, blockage anticipation, beam adaptation, sensing-assisted communication, UAV control, and dynamic resource allocation are representative examples because their decisions depend on latent system evolution or the consequences of candidate actions. The additional value of predictive representation learning decreases when system behavior is dominated by weakly predictable instantaneous variations, or when the relevant state is directly available and supports a compact, well-characterized decision model.

Effective CommJEPA design therefore depends on the alignment of predictive structure, decision-relevant information, and the cost of acquiring and using predictive information. This alignment defines the communication-aware design space developed in Section~III-B.

\section{The Mutual Evolution Between JEPA and Wireless Intelligence}
CommJEPA is motivated by the interaction between predictive representation learning and wireless systems. JEPA provides reusable latent representations for anticipating system evolution, integrating heterogeneous observations, and supporting downstream decisions. Wireless communications, in turn, introduces propagation physics, partial observability, multiple time scales, heterogeneous modalities, action-dependent dynamics, and deployment constraints. These characteristics shape the observations, prediction targets, and conditioning information used by predictive models, as well as their practical implementation.
\subsection{What JEPA Brings to Wireless Intelligence}
JEPA learns latent structure by predicting target representations from available observations. In wireless systems, observations and prediction targets may span time, space, network entities, and sensing modalities. This learning principle offers several capabilities relevant to wireless intelligence.
\begin{itemize}
\item \textbf{Reusable representations across wireless functions.}
Channel prediction, beam management, blockage detection, localization, mobility adaptation, and resource allocation often depend on shared latent factors, including propagation geometry, spatial relationships, mobility patterns, and network states. JEPA pretraining can encode these factors in a common representation that downstream tasks reuse through task-specific prediction heads or fine-tuning.
\item \textbf{Anticipatory inference of evolving wireless states.}
JEPA can predict future-relevant latent structure associated with channel conditions, blockage, mobility, interference, and traffic. These predictions can inform proactive beam adjustment, handover preparation, resource scheduling, and network adaptation. Their usefulness depends on whether the prediction horizon matches the response time of the target communication function.
\item \textbf{Cross-modal predictive learning.}
Wireless intelligence increasingly combines radio signals with sensing measurements, positioning information, visual observations, and network-side states. JEPA can learn predictive relationships across these sources, capturing shared environmental structure while retaining complementary modality-specific information. This is particularly useful when individual observations are incomplete, delayed, or unavailable.
\item \textbf{Representation learning from high-dimensional wireless observations.}
Wireless measurements contain complex variations caused by fading, noise, mobility, and hardware characteristics. Latent prediction can prioritize predictable, communication-relevant structure rather than reconstruct every raw observation. With suitable pretraining data, the resulting representations can also reduce dependence on task-specific annotations.
\item \textbf{Predictive support for optimization and control.}
Latent representations can connect wireless observations with downstream decision-making modules. When candidate actions are included as conditioning information, predictive models can estimate latent outcomes under different RIS configurations, UAV positions, beam selections, or resource-allocation decisions. These predictions can inform optimization, reinforcement learning, and control according to the relevant communication objectives.
\end{itemize}

The contribution of JEPA to wireless intelligence depends on the task and operating conditions. Propagation characteristics determine which dependencies are worth learning, while incomplete observations complicate state estimation and system dynamics determine suitable prediction horizons. Since communication actions can alter future network states, predictive models may also need to account for candidate actions. Practical deployment further depends on model complexity and inference cost, which influence the choice of prediction targets, conditioning information, latent representations, and evaluation metrics.

\subsection{How Wireless Communications Reshapes JEPA}
Wireless communications reshapes the predictive learning problem itself. A JEPA model must account for the physical dependencies underlying wireless observations, hidden states that cannot be directly observed, different time scales of system evolution, and the effects of communication actions. Its representations must also remain useful under practical resource constraints. In CommJEPA, these requirements translate into six design considerations:
\begin{itemize}
\item \textbf{Physics-aware predictive structure:} Predictive representations and targets should account for propagation geometry, antenna relationships, mobility, blockage, and other physical dependencies governing wireless system evolution.
\item \textbf{Latent-state inference under partial observability:} Predictive representations should integrate incomplete or delayed observations to infer communication-relevant hidden states and characterize uncertainty when multiple latent states remain plausible.
\item \textbf{Time-scale-dependent predictability:} Observation-history length, prediction horizon, and target definition should match the temporal scales of the underlying processes, since channel variation, mobility, blockage, traffic, and topology exhibit different predictive horizons.
\item \textbf{Multimodal predictive learning:} Representations should capture shared structure across radio, sensing, positioning, visual, and network observations while preserving complementary information specific to individual modalities.
\item \textbf{Action-dependent prediction:} Candidate communication and control actions can serve as conditioning information because RIS configuration, UAV movement, beam selection, and resource allocation can alter subsequent system states.
\item \textbf{Resource-aware predictive abstraction:} Latent dimensionality, predictor complexity, inference frequency, and information exchange should reflect communication and computation budgets, balancing predictive benefits against the costs of acquiring and using information.
\end{itemize}
Table~\ref{tab:commjepa_design} summarizes how these wireless characteristics translate into CommJEPA design implications and predictive insights. 
\begin{table*}[!t]
\centering
\caption{Wireless characteristics, predictive insights, and corresponding CommJEPA design implications.}
\label{tab:commjepa_design}

\renewcommand{\arraystretch}{1.2}
\setlength{\tabcolsep}{4pt}
\footnotesize

\begin{tabularx}{\textwidth}{
>{\raggedright\arraybackslash\columncolor{wirelessblue}}p{0.17\textwidth}
>{\raggedright\arraybackslash\columncolor{jepabeige}}X
>{\raggedright\arraybackslash\columncolor{jepabeige}}p{0.25\textwidth}
>{\raggedright\arraybackslash\columncolor{intelligencegreen}}p{0.18\textwidth}
}
\toprule

\textbf{Wireless Property}
&
\textbf{CommJEPA Design Implication}
&
\textbf{Predictive Insight}
&
\textbf{Example Scenario}
\\

\midrule

Physical propagation structure
&
Account for geometry, propagation knowledge, and physical consistency when selecting observations and designing prediction targets
&
Physically meaningful and transferable structure beyond signal-level correlations
&
Channel evolution and blockage-aware modeling
\\

\specialrule{0.25pt}{0pt}{0pt}

Partial observability
&
Aggregate temporal history and infer latent states from incomplete, noisy, or delayed observations
&
Prediction should recover decision-relevant hidden states and account for ambiguity caused by incomplete observations
&
Blockage inference and channel tracking
\\

\specialrule{0.25pt}{0pt}{0pt}

Multi-timescale dynamics
&
Match observation-history length and prediction horizon to the temporal scales of wireless dynamics
&
Time-scale-dependent predictability means that different horizons expose different predictable structures in channel, mobility, traffic, and topology
&
Beam tracking and mobility management
\\

\specialrule{0.25pt}{0pt}{0pt}

Heterogeneous observations
&
Align shared cross-modal structure while preserving complementary and modality-specific information
&
Predictability may arise from shared latent factors across modalities as well as complementary cues unavailable from a single observation source
&
Sensing-assisted communication
\\

\specialrule{0.25pt}{0pt}{0pt}

Action-dependent evolution
&
Condition latent prediction on candidate communication or control actions
&
Action-dependent prediction captures how future wireless conditions change under intervention and supports decision-aware prediction
&
RIS reconfiguration, UAV positioning, and resource control
\\

\specialrule{0.25pt}{0pt}{0pt}

Resource constraints
&
Adapt latent dimensionality, prediction complexity, and information exchange to communication and computation budgets
&
Balance predictive benefits against information acquisition, processing, and exchange costs
&
Distributed edge agents
\\

\bottomrule
\end{tabularx}
\end{table*}

These considerations are coupled rather than independent. In wireless systems, propagation structure, observation mechanisms, temporal dynamics, available modalities, candidate actions, and resource constraints jointly determine what can be inferred and what is useful to predict. Temporal history and complementary observations can help recover communication-relevant hidden states, while the appropriate prediction horizon depends on the dynamics associated with the intended decision. When actions influence subsequent connectivity, interference, or resource availability, action-conditioned prediction further distinguishes possible system evolutions under different interventions. At the same time, latency, energy, computation, memory, and signaling budgets constrain the amount and form of predictive information that can be represented and exchanged.
JEPA provides the general principle of latent prediction, while wireless communications determines which predictive relationships are physically meaningful, inferable from available observations, relevant to communication decisions, and feasible to deploy. CommJEPA incorporates these requirements into predictive representation learning itself, allowing the latent representation, prediction target, conditioning mechanism, and prediction horizon to be shaped by the structure and operational requirements of the wireless system.

\section{From Predictive Representations to Wireless Decisions}

The value of CommJEPA becomes clear when predictive representations support wireless decision-making. Wireless conditions evolve with channel dynamics, mobility, blockage, interference, and network load, while communication and control actions can further affect subsequent system states. Future-relevant representations can therefore provide useful information for proactive adaptation and optimization.

\begin{figure*}[t]
\centering
\includegraphics[width=\textwidth]{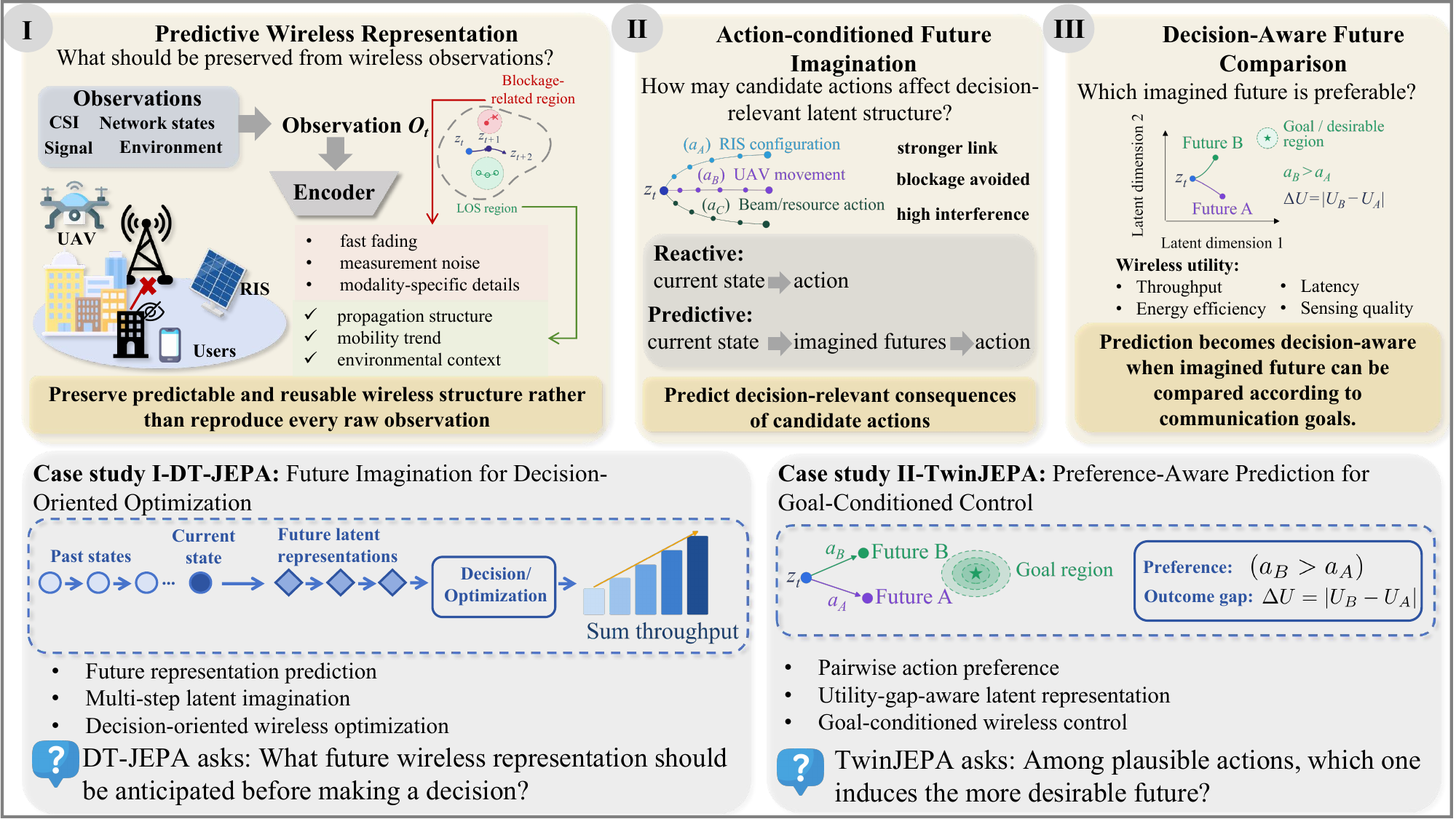}
\caption{From predictive representations to wireless decision-making in CommJEPA. Predictive representation learning captures future-relevant wireless structure and exposes decision-relevant information to downstream optimization or control. DT-JEPA illustrates future-state prediction in a dynamic wireless environment, while TwinJEPA illustrates objective-conditioned action-preference prediction.}
\label{fig:commjepa_decision}
\end{figure*}

The information learned and predicted by CommJEPA depends on the available observations, prediction targets, conditioning information, training objective, and prediction horizon. These choices should reflect the requirements of each communication task. Short-term channel estimation may require fine-grained signal features and rapid fading dynamics, whereas mobility management and blockage anticipation may rely more on persistent temporal and environmental patterns. When communication actions affect future network states, action-conditioned prediction can help evaluate candidate decisions according to their expected communication outcomes. As illustrated in Fig.~\ref{fig:commjepa_decision}, CommJEPA connects predictive representation learning with downstream optimization or control. The learned representation can characterize expected system evolution or the future consequences associated with candidate actions, while downstream modules use this information according to objectives such as throughput, energy efficiency, and resource utilization. This separation preserves flexibility in the choice of decision mechanism and enables predictive information to support communication decisions.

The following two examples illustrate these complementary pathways. DT-JEPA predicts future latent states in a dynamic wireless environment and shows how representations aligned with system evolution can support downstream optimization. TwinJEPA learns action preferences under different communication objectives and shows how goal-conditioned representations can support communication and computation decisions. The two cases therefore provide concrete realizations of the CommJEPA principles developed in the preceding sections.

\subsection{Future-State Prediction with DT-JEPA}

DT-JEPA illustrates the future-state prediction pathway of CommJEPA by using predicted latent wireless states to support decision-oriented optimization~\cite{ni2026dtjepalearningpredictfuture}. This example reflects the principles of anticipatory inference and prediction-target design aligned with wireless dynamics. By learning representations of future system conditions, DT-JEPA provides downstream decision modules with information about expected system evolution. We consider a UAV-assisted mobile edge computing (MEC) scenario in which a UAV provides computation offloading services to 15 mobile ground users within a \(1000\times600~\mathrm{m}^2\) area. The UAV moves among 25 candidate hovering locations, while user mobility follows a Gauss--Markov model with an average speed of \(5~\mathrm{m/s}\). The system state contains the locations of the UAV and ground users, and each action jointly selects the user to be served and the UAV's next hovering location. In this setting, user mobility continually changes the communication and computation geometry. The prediction target therefore needs to reflect the evolution of the wireless environment over the relevant decision horizon. DT-JEPA captures this evolving structure through future latent-state prediction, allowing the downstream decision process to incorporate anticipated system conditions when selecting users and UAV locations.

\begin{figure}[t]
\centering
\includegraphics[width=0.8\linewidth]{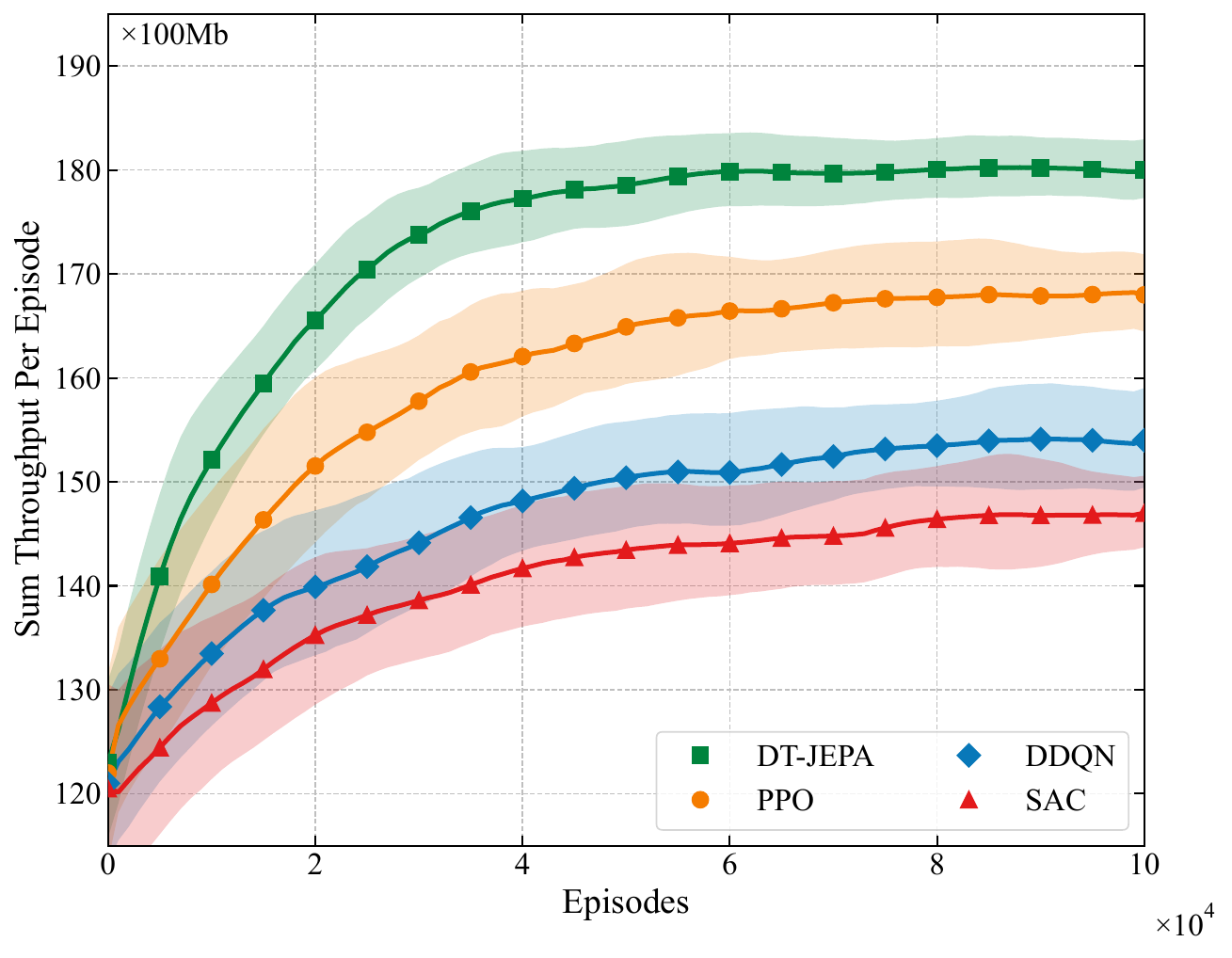}
\caption{DT-JEPA for future-state prediction in UAV-assisted MEC. Predicted latent representations capture future-relevant system evolution and provide information for downstream optimization of user selection and UAV placement.}
\label{fig:dtjepa}
\end{figure}

Figure~\ref{fig:dtjepa} compares DT-JEPA with PPO, DDQN, and SAC in terms of sum throughput during training. In the reported setting, DT-JEPA achieves a higher throughput level and converges more rapidly than the compared baselines. These results connect latent prediction with communication utility. The learned representation captures future-relevant system information, while the downstream decision mechanism uses that information to improve wireless resource utilization. This case illustrates two important CommJEPA design principles. First, the prediction target reflects the temporal evolution of a mobile wireless system, demonstrating the importance of aligning predictive representations with the dynamics relevant to downstream decisions. Second, the representation is evaluated through its contribution to an operational communication objective, connecting latent prediction with measurable system-level performance.

\subsection{Action-Preference Prediction with TwinJEPA}

Future-state prediction describes how the wireless environment may evolve. A complementary decision problem concerns the relative desirability of candidate actions under a specified communication objective. In this setting, predictive representations should preserve information that distinguishes the expected consequences of alternative decisions.

TwinJEPA explores this direction through temporal action-preference learning for goal-conditioned control~\cite{you2026twinjepa}. It connects to the CommJEPA principles of action-dependent prediction and objective-aware representation learning. The model learns predictive representations that encode distinctions among candidate actions according to their downstream consequences and the goal being pursued. To illustrate this approach, we consider a multi-user MEC scenario with three ground users offloading computation to a fixed edge server over temporally correlated Rayleigh fading channels. The users share a transmit-power budget, and each action jointly determines the power allocation and computation-offloading ratio of each user. These decisions lead to different energy and rate consequences. This setting highlights the role of objective-conditioned prediction. The same system state can support different trade-offs between energy consumption and communication rate, and the preferred action therefore depends on the communication objective. A useful predictive representation should preserve information relevant to these trade-offs so that candidate actions can be compared according to the objective governing the decision.

\begin{figure}[t]
\centering
\includegraphics[width=\linewidth]{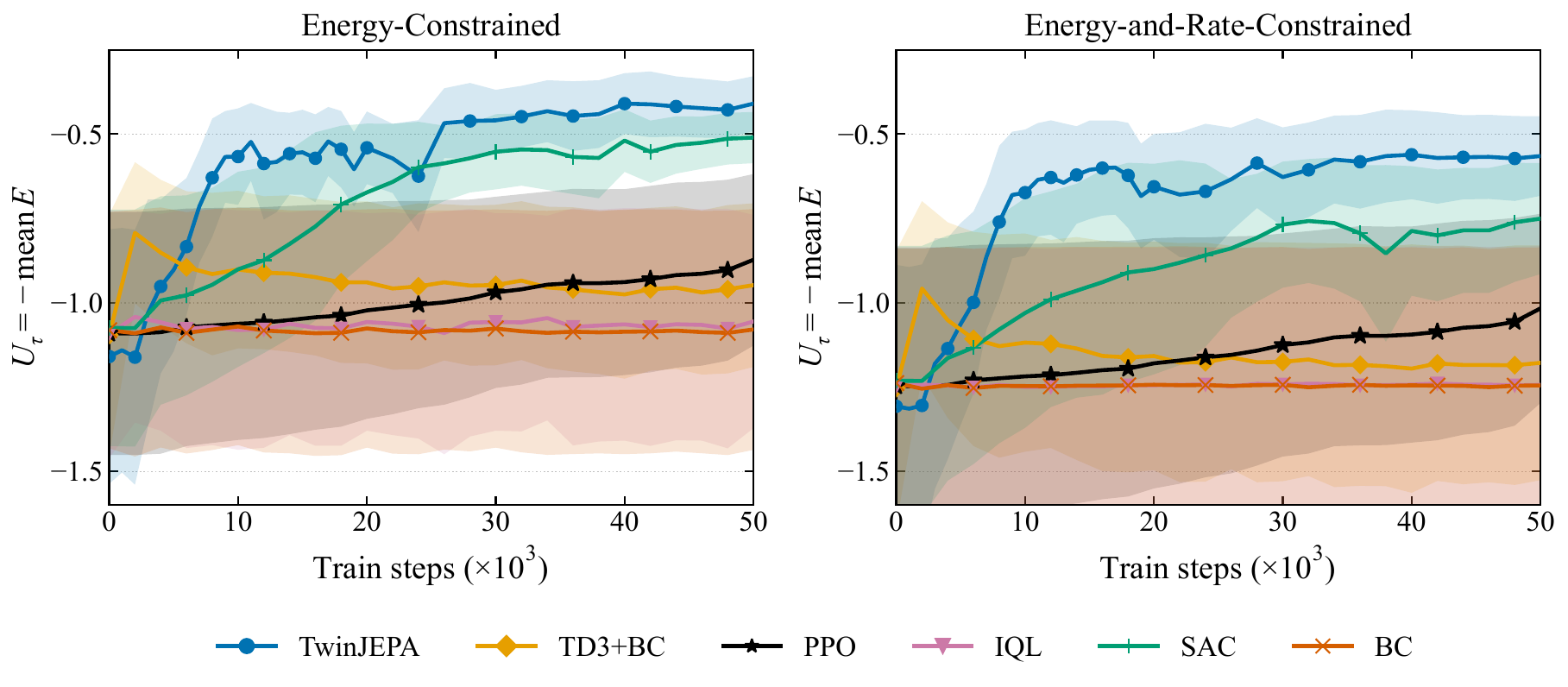}
\caption{TwinJEPA instantiated for preference-aware control in multi-user uplink MEC. Goal-conditioned predictive representations distinguish candidate actions according to their energy and rate consequences under different communication objectives.}
\label{fig:twinjepa}
\end{figure}

As shown in Fig.~\ref{fig:twinjepa}, TwinJEPA is evaluated under Energy-Constrained and Energy-and-Rate-Constrained settings. The two settings reflect different trade-offs between energy consumption and communication rate. Across both objectives, TwinJEPA achieves favorable utility compared with the considered learning-based baselines, illustrating how goal-conditioned predictive representations can support preference-aware wireless control.

\begin{figure}[t]
\centering
\includegraphics[width=\linewidth]{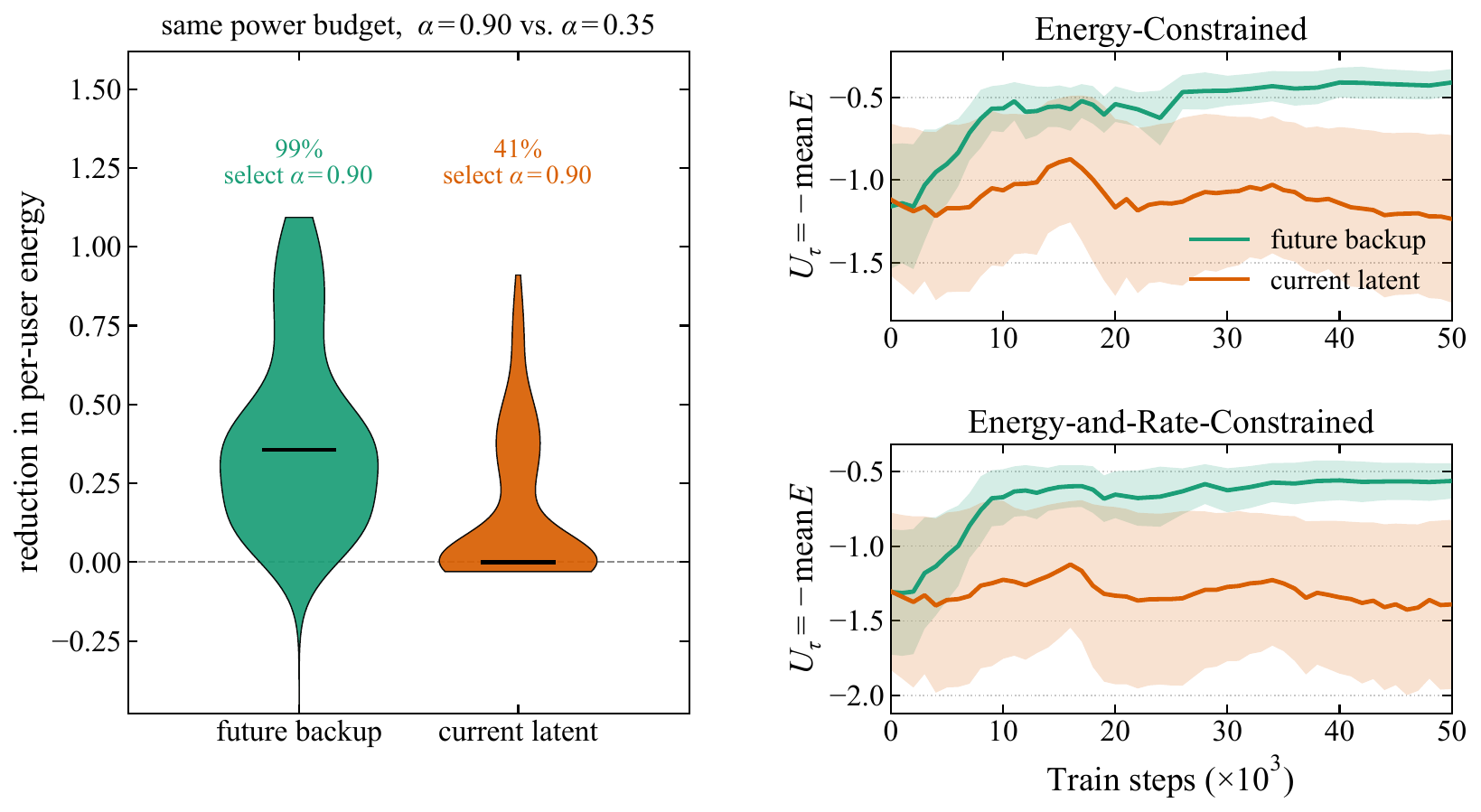}
\caption{Role of future-oriented predictive learning in TwinJEPA. The left panel illustrates how the future-backup target affects the preference between alternative offloading decisions, while the right panels show the corresponding control utility under the two communication objectives.}
\label{fig:td-target}
\end{figure}

Beyond the overall control performance, an important question is what future-oriented predictive learning contributes to the learned action preference. Fig.~\ref{fig:td-target} provides a complementary view by comparing the original future-backup target with a variant that removes the future-oriented component. With future backup, TwinJEPA exhibits a clearer preference between alternative offloading decisions and stronger control performance under both communication objectives. When this component is removed, the action preference becomes less distinct and the corresponding utility deteriorates.

This comparison provides an intuitive view of the role of predictive learning in TwinJEPA. Rather than relying only on representations of the current system state, the future-oriented target encourages the learned representation to retain information about how candidate actions may lead to different communication and computation consequences. The resulting preference structure can therefore reflect decision-relevant differences that are not apparent from the current observation alone. The comparison also complements the overall TwinJEPA results by showing that weakening future-oriented prediction reduces both the distinction between candidate actions and the resulting control performance. 

\subsection{What the Two Examples Reveal About CommJEPA}
DT-JEPA and TwinJEPA illustrate complementary connections between predictive representation learning and wireless decision-making. Their relationships to the CommJEPA design principles discussed earlier can be understood from three aspects.
\begin{itemize}
\item \textbf{Future-state prediction and anticipatory inference.}
DT-JEPA illustrates how latent representations of future wireless states capture the effects of evolving mobility and communication geometry. In UAV-assisted MEC, this predictive target reflects the temporal variations of the wireless environment that influence subsequent resource allocation. Rather than relying solely on the current system state, the learned representation provides information about its potential evolution, connecting predictive target design with the temporal dynamics relevant to downstream optimization.

\item \textbf{Action preference and objective-conditioned prediction.}
TwinJEPA extends predictive representation learning from anticipating system evolution to distinguishing the potential consequences of alternative actions. Its goal-conditioned formulation incorporates candidate actions and communication objectives into the prediction process, allowing the learned representations to reflect different energy and rate requirements. The comparison with the current-latent target further suggests that future-oriented prediction helps preserve action-dependent distinctions, while objective conditioning determines which aspects of the predicted consequences are relevant to the decision. This illustrates the complementary roles of predictive targets and conditioning information in learning action preferences.
\item \textbf{Predictive representations and downstream decision mechanisms.}
The two examples also illustrate different ways of connecting predictive representations with wireless decisions. In DT-JEPA, anticipated future-state information is used by downstream optimization to account for changes in the wireless environment. In TwinJEPA, action-dependent representations provide information for comparing candidate decisions under specified communication objectives. In both cases, predictive learning organizes decision-relevant information at the representation level, while the subsequent optimization or control mechanism determines how this information is used to meet operational requirements.
\end{itemize}

Across these examples, two complementary pathways emerge for applying JEPA-based predictive learning to wireless intelligence. DT-JEPA organizes prediction around future system states and their temporal evolution, whereas TwinJEPA focuses on action-dependent consequences conditioned on candidate actions and communication objectives. The distinction lies not only in the prediction targets but also in the conditioning mechanisms that determine what information is relevant to subsequent decisions. Other principles discussed earlier, including cross-modal predictive organization, physical consistency, and resource-aware representation design, further extend this design space to broader wireless settings.

\section{Open Challenges and Future Directions}

CommJEPA aims to move wireless AI from reactive processing toward predictive and interactive intelligence. Realizing this vision requires predictive representations that remain physically meaningful, robust to incomplete observations, transferable across environments, useful for decision-making, and practical under communication and computation constraints. Several challenges remain open.

\subsection{Physics-Grounded Predictive Representations}

Wireless observations are shaped by propagation geometry, multipath, mobility, blockage, antenna configurations, and hardware characteristics. Data-driven latent prediction may overlook important physical dependencies, while strong analytical constraints can reduce adaptability to changing wireless environments. Future CommJEPA designs should incorporate lightweight physical inductive biases through prediction-target design, conditioning information, and physical consistency constraints. An important research direction is to develop predictive representations that preserve relevant physical relationships while adapting to diverse wireless conditions.

\subsection{Multimodal Prediction under Incomplete Observations}

Future wireless systems will increasingly combine radio, sensing, positioning, visual, mobility, and network-side information. These observations may be asynchronous, noisy, incomplete, or unavailable at inference time. CommJEPA should therefore preserve shared and modality-specific information while remaining robust to missing observations. Uncertainty-aware prediction is also important when the available observations are consistent with multiple plausible future latent states.

\subsection{Interactive Prediction and Wireless World Models}

Wireless communication systems are inherently interactive because communication and control actions can influence subsequent system conditions. CommJEPA may therefore extend beyond passive state prediction by modeling decision-relevant consequences of candidate actions. Important questions include how actions should condition latent prediction, how uncertainty should be handled over longer horizons, and how environmental dynamics can be distinguished from action-induced changes. Addressing these issues could support compact wireless world models that inform optimization, reinforcement learning, and control without replacing the downstream decision mechanism.

\subsection{Generalization, Adaptation, and Representation Reuse}

A useful CommJEPA representation should transfer across locations, frequencies, antenna configurations, mobility patterns, hardware platforms, and network conditions. Future work should identify which latent structures are broadly reusable and which remain environment specific. Evaluation should therefore extend beyond single-task accuracy to include transfer performance, adaptation efficiency, robustness under domain shift, and reuse across multiple wireless objectives.

\subsection{Resource-Efficient and Distributed CommJEPA}

Practical deployment also requires predictive learning under limited computation, memory, energy, latency, and signaling budgets. Compact latent spaces, lightweight predictors, distributed encoding, selective information exchange, and adaptive inference are therefore important design directions. More broadly, CommJEPA should consider not only prediction quality but also the cost of producing and exchanging predictive information. This calls for joint evaluation of accuracy, representation size, communication overhead, inference latency, energy consumption, and downstream decision benefit.

Overall, future CommJEPA research should move beyond demonstrating latent prediction on individual wireless tasks. The broader goal is to develop predictive representations that are physically grounded, uncertainty aware, transferable, interaction aware, and efficient enough for real wireless systems. Such representations could provide a reusable interface between wireless observations and intelligent network decisions.

\section{Conclusion}

Future wireless intelligence requires models that capture predictable and reusable structure as communication environments evolve. In this article, we introduced \emph{Communication JEPA} (\emph{CommJEPA}) as a communication-oriented framework for predictive representation learning that accommodates different neural architectures and control algorithms. Its central premise is the bidirectional relationship between JEPA and wireless communications. JEPA enables wireless systems to learn anticipatory latent representations, while the physical structure of wireless environments, incomplete observations, heterogeneous information sources, interactive dynamics, and deployment constraints shape the design of predictive learning. These factors influence the information to be represented, the targets to be predicted, and the conditions that guide prediction.
The DT-JEPA and TwinJEPA case studies demonstrate two complementary ways in which predictive representations can support wireless decision-making through future latent-state prediction and decision-relevant action preferences. These cases represent selected instantiations of CommJEPA and illustrate its potential across different wireless tasks. More broadly, CommJEPA offers a foundation for developing wireless AI that anticipates relevant system evolution and supports adaptation and decision-making. Further research is needed to establish predictive representations that generalize across environments, accommodate heterogeneous observations, and meet practical deployment requirements.

\bibliographystyle{IEEEtran}
\bibliography{Ref}

@article{you2026twinjepa,
  title={TwinJEPA: Action-Preferred Predictive Representations for Goal-Conditioned Control},
  author={You, Feiran and Du, Hongyang},
  journal={arXiv preprint arXiv:2610.02922},
  year={2026}
}

@article{yang2025decision,
  title={Decision-making large language model for wireless communication: A comprehensive survey on key techniques},
  author={Yang, Ning and Fan, Mingrui and Wang, Wentao and Zhang, Haijun},
  journal={IEEE Communications Surveys \& Tutorials},
  year={2025},
  publisher={IEEE}
}

@ARTICLE{11155147,
  author={Zhang, Ping and Niu, Kai and Wang, Xiaoyun and Liu, Yiming and Liang, Zijian and Dong, Chen and Dai, Jincheng and Xu, Xiaodong and Xu, Wenjun and Zhang, Zhi and Wang, Guangyu and Li, Yanlu and Wu, Di and Wu, Hequan},
  journal={IEEE Communications Surveys \& Tutorials}, 
  title={ComAI: The Convergence of Communication and Artificial Intelligence}, 
  year={2026},
  volume={28},
  number={},
  pages={2163-2197},
  doi={10.1109/COMST.2025.3608174}}

@article{lecun2022path,
  title={A path towards autonomous machine intelligence version 0.9. 2, 2022-06-27},
  author={LeCun, Yann and others},
  journal={Open Review},
  volume={62},
  number={1},
  pages={1--62},
  year={2022}
}

@inproceedings{assran2023self,
  title={Self-supervised learning from images with a joint-embedding predictive architecture},
  author={Assran, Mahmoud and Duval, Quentin and Misra, Ishan and Bojanowski, Piotr and Vincent, Pascal and Rabbat, Michael and LeCun, Yann and Ballas, Nicolas},
  booktitle={2023 IEEE/CVF Conference on Computer Vision and Pattern Recognition (CVPR)},
  pages={15619--15629},
  year={2023},
  organization={IEEE}
}

@article{jiang2026large,
  title={From large ai models to agentic ai: A tutorial on future intelligent communications},
  author={Jiang, Feibo and Pan, Cunhua and Wang, Kezhi and Michiardi, Pietro and Dobre, Octavia A and Debbah, Merouane},
  journal={IEEE Journal on Selected Areas in Communications},
  year={2026},
  publisher={IEEE}
}

@ARTICLE{10930485,
  author={Pivoto, Diego Gabriel Soares and Figueiredo, Felipe A. P. de and Cavdar, Cicek and Tejerina, Gustavo Rodrigues de Lima and Mendes, Luciano Leonel},
  journal={IEEE Communications Surveys \& Tutorials}, 
  title={A Comprehensive Survey of Machine Learning Applied to Resource Allocation in Wireless Communications}, 
  year={2026},
  volume={28},
  number={},
  pages={1986-2053}
  }

@ARTICLE{11611638,
  author={Shah, Syed Danial Ali and Nezami, Zeinab and Gregory, Mark A.},
  journal={IEEE Wireless Communications}, 
  title={Agentified 6G Edge Intelligence: Exploring Embodied AI in Open RAN and Future Research Directions}, 
  year={2026},
  volume={},
  number={},
  pages={1-9}}

@ARTICLE{11316403,
  author={Zhang, Han and Farzanullah, Mohammad and Ghassemi, Mohammad and Bin Sediq, Akram and Afana, Ali and Erol-Kantarci, Melike},
  journal={IEEE Communications Surveys \& Tutorials}, 
  title={Multi-Modal Data-Enhanced Foundation Models for Prediction and Control in Wireless Networks: A Survey}, 
  year={2026},
  volume={28},
  number={},
  pages={4359-4393}}

@ARTICLE{11248867,
  author={Ríos-Guiral, Santiago and Lahmadi, Abdelkader and Botero, Juan F. and Gutiérrez, Sergio A.},
  journal={IEEE Communications Surveys \& Tutorials}, 
  title={Leveraging Reinforcement Learning for Traffic Engineering in Programmable Networks: A Survey}, 
  year={2026},
  volume={28},
  number={},
  pages={1318-1350}}

@misc{chu2026wirelessjepamultiantennafoundationmodel,
      title={WirelessJEPA: A Multi-Antenna Foundation Model using Spatio-temporal Wireless Latent Predictions}, 
      author={Viet Chu and Omar Mashaal and Hatem Abou-Zeid},
      year={2026},
      eprint={2601.20190},
      archivePrefix={arXiv},
      primaryClass={eess.SP},
      url={https://arxiv.org/abs/2601.20190}, 
}

@misc{ni2026dtjepalearningpredictfuture,
      title={DT-JEPA: Learning to Predict Future Representations for Decision-Making}, 
      author={Changxu Ni and Yiyang Ni and Jun Li and Feiran You and Hongyang Du and Haitao Zhao and Shi Jin},
      year={2026},
      eprint={2610.32657},
      archivePrefix={arXiv},
      primaryClass={cs.LG},
      url={https://arxiv.org/abs/2610.32657}, 
}

\end{document}